\documentclass{aastex701}

\usepackage{bm}
\usepackage{amsmath}
\usepackage{comment}

\graphicspath{{./}{figures/}}

\begin{document}

\title{On the Reparameterization Between Cartesian Position--Velocity Vectors and Orbital Elements in the Kepler Problem}

\author[0000-0003-1298-9699]{Kento Masuda}
\affiliation{Department of Earth and Space Science, Graduate School of Science, Osaka University, Toyonaka, Osaka 560-0043, Japan}
\email{kmasuda@ess.sci.osaka-u.ac.jp}

\author[0000-0003-1298-9699]{Kansuke Nunota}
\affiliation{Department of Earth and Space Science, Graduate School of Science, Osaka University, Toyonaka, Osaka 560-0043, Japan}
\email{}


\begin{abstract}

Reparameterization from the standard set of orbital elements to Cartesian position--velocity vectors can be computationally advantageous for orbit inference problems, particularly when orbital elements are weakly constrained. 
Here we present compact analytic expressions for the Jacobian determinants of this transformation and its variants, which enable consistent transformation of prior probability densities under reparameterization and are therefore useful for a Bayesian treatment of such problems.
We then use these results to clarify the application of this reparameterization in microlensing and astrometric contexts.
We first revisit the widely used formulation of lens orbital motion during binary microlensing events presented by \citet{2011ApJ...738...87S}. We show that their parameterization inadvertently adopts an incorrect definition of the longitude of the ascending node with respect to the sky-projected binary axis at a reference epoch, which renders the intermediate Jacobian formally singular. Using our closed-form expression, we provide a corrected analytic derivation of the Jacobian for this transformation and show that the resulting formula remains effectively unchanged when the longitude of the ascending node is properly defined with respect to an axis independent of the binary orbit.
We also perform an explicit quantitative comparison of astrometric orbit fitting using a gradient-based Markov Chain Monte Carlo algorithm under the two parameterizations, and find that reparameterizing to Cartesian state vectors improves sampling efficiency and robustness relative to orbital-element sampling.

\end{abstract}




\newcommand{\x}{\bm{x}}
\newcommand{\xv}{(\bm{x}, \bm{v})}
\renewcommand{\v}{\bm{v}}
\newcommand{\elm}{\bm{\theta}}
\newcommand{\jkep}{J_\mathrm{Kep}}
\newcommand{\jkepfive}{J'_\mathrm{Kep}}
\newcommand{\jkepfour}{J^{(4)}_\mathrm{Kep}}
\newcommand{\tref}{t_\mathrm{ref}}
\newcommand{\re}{r_\mathrm{E}}
\newcommand{\gammav}{\bm{\gamma}}
\newcommand{\omeganode}{\Omega_\mathrm{node}}

\section{Introduction}

A bound orbit in the Kepler problem is an ellipse and is conveniently characterized by orbital elements, 
which are functions of the constants of motion together with the orbital phase and provide a compact description of the size, shape, and orientation of the orbit \citep[e.g.,][]{2014grav.book.....P}. 
On the other hand, observables of stellar binaries and planetary systems are often more directly related to the time-dependent position $\x(t)$ and velocity $\v(t)$, and so the conversion from the elements $\elm$ to ($\x(t)$, $\v(t)$) plays a central role in modeling actual observations.
This conversion has long been established and is described in many standard textbooks \citep[e.g.,][]{1999ssd..book.....M}.

The connection between observables and $(\x(t),\v(t))$ has motivated orbit-inference methods that use the Cartesian state vectors {\it at a chosen reference epoch} $(\x,\v)$ rather than orbital elements as sampling or optimization parameters, especially when the orbit is only weakly constrained \citep[e.g.,][]{2011ApJ...738...87S, 2021RNAAS...5..162F}.\footnote{This reparameterization is expected to be most useful when the orbit is weakly constrained and the posterior geometry in orbital elements is highly non-Gaussian or strongly correlated; if the posterior is already compact and well behaved in orbital elements, the improvement is expected to be modest.}
When such a reparameterization is used in Bayesian inference, the one-to-one mapping between the orbital elements and state vectors alone is not sufficient. 
Because Bayesian analyses involve prior probability density functions (PDFs) for the parameters in addition to the likelihood function, one also needs to transform PDFs:
\begin{align}
    p(\x, \v)d\x d\v = p(\elm)d\elm.
\end{align}
This also requires the determinant of the $6\times 6$ Jacobian matrix:
\begin{align}
\label{eq:jkep}
    \jkep \equiv {\partial {\xv} \over \partial \elm}.
\end{align}
For example, if we adopt a uniform prior for $\xv$, $p_{\mathcal{U}\xv}=\mathrm{const.}$, the corresponding prior PDF for $\elm$ is given by
\begin{align}
    p_{\mathcal{U}\xv}(\elm) = p_{\mathcal{U}\xv}(\x, \v)\left| \det \jkep \right|\propto \left|\det \jkep(\elm)\right|,
\end{align}
where the last expression indicates that $\jkep$ is to be evaluated at $\xv = (\x(\elm), \v(\elm))$.
Alternatively, if we want a uniform prior for $\elm$, $p_{\mathcal{U}(\elm)}=\mathrm{const.}$, the prior PDF for $\xv$ should be
\begin{align}
\label{eq:pxv_from_pelm}
    p_{\mathcal{U}(\elm)}(\x, \v) = {1 \over |\det \jkep|}\,p_{\mathcal{U}(\elm)}(\elm) \propto {1 \over |\det \jkep(\x, \v)|}.
\end{align}

This prior transformation is sometimes left implicit. For example, \citet{2021RNAAS...5..162F} compared the efficiency of orbit fitting using standard sets of orbital elements and Cartesian state vectors, but did not explicitly account for the different priors induced by this transformation. As a result, the two parameterizations corresponded to different posterior distributions, making it difficult to isolate the effect of the reparameterization itself (see Section~\ref{sec:astrometry} for more details).
When treated explicitly, the Jacobian determinant has been evaluated numerically in the exoplanet literature --- e.g., by calculating the elements of $\jkep$ using the well-known relation between $\xv$ and elements \citep[e.g.,][]{2011ApJ...738...87S}, or by using automatic differentiation \citep[e.g.,][]{2023AJ....166..164T} without deriving or implementing an analytic expression.

Although the determinant of this transformation has a simple analytic form that follows directly from the canonical phase-space volume element in Delaunay variables and has appeared in stellar-dynamical contexts \citep[e.g.,][]{2003ApJ...599..237P,2020MNRAS.491.1941T}, it does not appear to be routinely used in this form as a Jacobian determinant for prior transformations in exoplanet orbit-inference applications.
In Section~\ref{sec:detjkep}, we therefore review this derivation and present compact closed-form expressions for the Jacobian determinants associated with this transformation and its useful variants. These expressions make the prior conversion transparent and easy to implement in Bayesian orbit inference.
As an illustrative example of such an application, we revisit the formulation of the lens orbital motion by \citet{2011ApJ...738...87S}, which has been widely used in microlensing analyses (Section~\ref{sec:skowron}). We point out a minor but conceptual error in their parameterization, present a modified derivation of their formula taking advantage of our analytic formulation, and explain why the result essentially remains {\it correct}, except that the definition of the longitude of the ascending node needs to be updated.
We also perform an explicit quantitative examination of the performance of such reparameterization for interpreting orbit data with incomplete phase coverage (Section~\ref{sec:astrometry}).

\section{Closed-Form Expressions for the Jacobian determinant}\label{sec:detjkep}

\subsection{The Full Kepler Problem}\label{ssec:detjkep6}

Consider the Kepler problem with given total mass $\mathcal{M}$.
Let $\x$ and $\v$ be the Cartesian position and velocity vectors at a reference epoch $\tref$, respectively.
Let $a$, $e$, $i$, $\Omega$, $\omega$, and $M$ be the semi-major axis, eccentricity, inclination, longitude of the ascending node, argument of periapsis referred to the ascending node, and mean anomaly at $\tref$, respectively.
Then the determinant of the Jacobian matrix for the transformation between $(\x,\v)$ and $(a,e,i,\Omega,\omega,M)$ is given by
\begin{align}
\label{eq:detjkep}
\det \jkep \equiv \det\left({\partial(\x,\v)\over\partial(a,e,i,\Omega,\omega,M)}
\right) = \frac{1}{2}\,\mu^{3/2}\, a^{1/2}\, e\, \sin i,
\end{align}
where $\mu\equiv \mathcal{G}\mathcal{M}$ with $\mathcal{G}$ being Newton's gravitational constant.
The $\sin i$ dependence --- along with the independence of $\Omega$ and $\omega$ --- implies that the orbits are isotropically oriented when the joint distribution of $\xv$ has a uniform density. The formula, in its present form, is valid only for bound orbits, for which $\x$ and $\v$ satisfy
\begin{align}
    {1\over 2}|\v|^2 - {\mu \over |\x|} < 0.
\end{align}
\\

\noindent\textbf{Derivation} --- The Delaunay variables defined as
\begin{align}
(l, g, h, L, G, H) = (M, \omega, \Omega, \sqrt{\mu a}, L\sqrt{1-e^2}, G\cos i)
\end{align}
form a set of canonical variables as are $\xv$ for the Hamiltonian
\begin{align}
    \mathcal{H}={1\over 2}|\v|^2 - {\mu \over |\x|} = -{\mu^2 \over 2L^2}
\end{align}
defined as the energy per reduced mass \citep[e.g.,][]{2023dyps.book.....T}.
The map between $\xv$ and $(l,g,h,L,G,H)$ is canonical, and we have 
\begin{align}
\det\left({\partial (\x, \v) \over \partial(l,g,h,L,G,H)}\right) = 1.
\end{align}
Therefore,
\begin{align}
\det J_{\rm Kep} 
= -\det \left({\partial(\x,\v) \over \partial(a,e,i,M,\omega,\Omega)}\right)
= -\det \left(
\frac{\partial(l,g,h,L,G,H)}
     {\partial(a,e,i,M,\omega,\Omega)}
\right)
= \det \left(
\frac{\partial(l,g,h,L,G,H)}
     {\partial(M,\omega,\Omega,a,e,i)}
\right)
=\det \left(
\frac{\partial(L,G,H)}
     {\partial(a,e,i)}
\right),
\end{align}
where the last equality holds because $(l,g,h)=(M,\omega,\Omega)$.\footnote{The sign of the determinant flips depending on column ordering. Here we defined $\det\jkep$ following the standard ordering of orbital elements, which is different from the natural ordering as implied by Delaunay elements. This results in sign flips in the determinant depending on whether the relevant permutation is even or odd, which also increases the risks of errors in the equations --- though in practice, only $|\det \jkep|$ matters and the sign is irrelevant.}
Furthermore, the last Jacobian is a lower triangular matrix:
\begin{align}
    \frac{\partial(L,G,H)}
     {\partial(a,e,i)}
     = \begin{pmatrix}
     {L \over 2a} & 0 & 0\\
     {L \over 2a}\sqrt{1-e^2} & -{Le \over \sqrt{1-e^2}} & 0 \\
     {L \over 2a}\sqrt{1-e^2}\cos i & -{Le \over \sqrt{1-e^2}}\cos i & -G\sin i
     \end{pmatrix},
\end{align}
and its determinant is obtained by multiplying the diagonal elements. 
Hence,
\begin{align}
\label{eq:jacobian}
\det J_{\rm Kep} 
=  {L^3 \over 2a}\, e\, \sin i
= \frac{1}{2}\,\mu^{3/2}\, a^{1/2}\, e\, \sin i.
\end{align}

\subsubsection{Expressions using Other Elements}

One may prefer to use the orbital period $P$ or mean motion $n=2\pi/P$ instead of $a$; or the time of periastron passage $\tau$ instead of $M=n(\tref-\tau)$. 
The following relations can be used to convert the argument of $\det\jkep$: 
\begin{align}
    {\partial M \over \partial \tau} &= -n = -\mu^{1/2}a^{-3/2},\\
    {\partial a \over \partial n} &=  -{2\over 3}{a \over n} = -{2\over 3}\mu^{1/3}n^{-5/3},\\
    {\partial a \over \partial P} &=  {2\over 3}{a \over P} = {2\over 3}(4\pi^2)^{-1/3}\mu^{1/3}P^{-1/3}.
\end{align}
For example, 
\begin{align}
\label{eq:detjkep_pm}
\det\jkep^{(PM)} \equiv  \det \left(\frac{\partial(\x,\v)}{\partial(P,e,i,\Omega,\omega,M)}\right)
&= {1\over 3}\mu^{3/2}a^{3/2}P^{-1}e\sin i = {\mu^2 \over 6\pi}\,e\sin i,\\
\label{eq:detjkep_atau}
\det\jkep^{(a\tau)} \equiv \det \left(\frac{\partial(\x,\v)}{\partial(a,e,i,\Omega,\omega,\tau)}\right)
&= -{1\over2}\,\mu a^{-1/2}e\sin i.
\end{align}

\subsubsection{Reduced Jacobian Eliminating the Nodal Degree of Freedom}\label{ssec:detjkep5}

In some practical applications, one overall rotation of the coordinate system can be chosen so that a particular coordinate takes a prescribed value. 
For example, in the microlensing application discussed in Section~\ref{sec:skowron}, one may choose $\Omega$ (or equivalently, the reference direction in the sky plane) such that the $y$-coordinate of the orbit vanishes at a given reference epoch. Imposing such a condition makes $\Omega$ an implicit function of the remaining orbital elements, and the transformation from orbital elements to Cartesian phase-space coordinates effectively becomes five-dimensional.

Here we mainly focus on the case in which the reference direction is chosen so that the $y$-coordinate is fixed. In this representation, it is natural to work with the reduced Jacobian
\begin{align}
    \jkepfive \equiv {\partial(x,z,v_x,v_y,v_z) \over \partial(a,e,i,\omega,M)},
\end{align}
which excludes both the redundant coordinate $y$ and the corresponding independent variable $\Omega$. An expression for $\det \jkepfive$ can be useful when one wishes to construct priors or coordinate parameterizations that reflect only the observable geometry, without carrying the extra degeneracy associated with $\Omega$. 
While $\det\jkep$ does not depend on $\Omega$, $\det\jkepfive$ is clearly different from $\det\jkep$: compared to $\jkep$, $\jkepfive$ removes the length $y$ from the numerator, and the angle $\Omega$ from the denominator, so its dimension differs from that of $\jkep$. We find that the full six-dimensional Jacobian factorizes as:
\begin{align}
\label{eq:detjkep5}
    \det\jkep = x \left.\det \jkepfive\right|_{y},
\end{align}
where the right-hand side of this equation is evaluated at a fixed $y$ as follows: Given $y$, the orientation of the $xy$-axes is fixed, and so the longitude of the ascending node is determined implicitly as $\Omega^*=\Omega(a,e,i,\omega,M,y)$. The coordinates $(x,z,v_x,v_y,v_z)$ are then evaluated using this $\Omega^*$. This means both $x$ and $\det\jkepfive$ depend on the chosen value of $y$, but their product does not, as is the case for the left-hand side.
\\

\noindent\textbf{Derivation} --- Because $\Omega$ represents rotation within the reference plane, $\x$ can be written as
\begin{align}
    \begin{pmatrix}
        x(\elm) \\ y(\elm) \\ z(\elm)
    \end{pmatrix} = \begin{pmatrix}
        \cos\Omega & -\sin\Omega & 0 \\
        \sin\Omega & \cos\Omega & 0 \\
        0 & 0 & 1
    \end{pmatrix}\,\begin{pmatrix}
        x_0(\elm') \\ y_0(\elm') \\ z_0(\elm')
    \end{pmatrix},
\end{align}
where $\x_0$ is the position vector that depends only on the elements other than $\Omega$, which we write as $\elm'$. This yields
\begin{align}
    {\partial y \over \partial \Omega} = x_0\cos\Omega - y_0\sin\Omega = x. 
\end{align}
Thus, 
\begin{align}
    \det\jkep 
    = \det\left({\partial(x,y,z,v_x,v_y,v_z)\over\partial(a,e,i,y,\omega,M)}\right)
    \cdot {\partial y \over \partial \Omega}
    = x\det\left({\partial(x,y,z,v_x,v_y,v_z)\over\partial(a,e,i,y,\omega,M)}\right)
    = x\det\left({\partial(x,z,v_x,v_y,v_z)\over\partial(a,e,i,\omega,M)}\right)
    = x\det\jkepfive.
\end{align}
From this derivation, it is straightforward to show that the pre-factor becomes $-y$ when $x$ is fixed instead. It is also possible to fix $v_x$ or $v_y$, in which case analogous relations hold.

\subsection{The 2D Kepler Problem}\label{ssec:detjkep4}

The orbits of transiting exoplanets have $i\approx \pi/2$ by construction. If we further set $i=\pi/2$ and $\Omega=0$ or $\pi$, the orbit is restricted to the $y=0$ plane, and the problem reduces to the 2D Kepler problem.\footnote{This reduction in dimensionality is qualitatively different from the one discussed in Section~\ref{ssec:detjkep5}, where one of the orbital elements is removed by expressing it as an implicit function of the remaining orbital elements. Here we are considering the case in which, for example, $i$ is physically fixed to $\pi/2$, rather than being solved for from the other elements.}
This is a common approximation when analyzing transit timing variations in multi-transiting systems \citep{2018haex.bookE...7A}, where the deviation from $i=\pi/2$ typically has a negligible impact on the observed transit times and the mutual orbital inclinations are often small. It is thus of interest to compute the Jacobian determinant in this case as well, although the merit of reparameterization using $\x$ and $\v$ has yet to be demonstrated for this class of problems. The result is:
\begin{align}
    \det\jkepfour \equiv \det\left({\partial(x, z, v_x, v_z)} \over \partial(a,e,\omega,M) \right) = {\mu\over 2}{e \over \sqrt{1-e^2}}.
\end{align}\\

\noindent\textbf{Derivation} --- The derivation follows that of the 3D Kepler problem. The Delaunay variables in the 2D Kepler problem are:
\begin{align}
    \left(l, g, L, G\right) = \left(M,\omega,\sqrt{\mu a}, L\sqrt{1-e^2}\right).
\end{align}
Thus,
\begin{align}
\notag
    \det\jkepfour &= \det\left(\partial(l,g,L,G) \over \partial(a,e,\omega,M)\right) = -\det\left(\partial(L,G)\over\partial(a,e)\right)\\
        &= -\det \begin{pmatrix}
        {1\over 2}{L\over a} & 0\\
        {1\over 2}{L\over a}\sqrt{1-e^2} & -{Le\over\sqrt{1-e^2}}
    \end{pmatrix}
    ={1\over 2}{L^2\over a}{e\over\sqrt{1-e^2}} = {\mu\over 2}{e \over \sqrt{1-e^2}}.
\end{align}


\subsection{Numerical Verification of the Formula}

We verified these expressions using functions $\x(\elm)$ and $\v(\elm)$ implemented in the {\tt jnkepler} package \citep{2024AJ....168..294M}. These functions are implemented in JAX \citep{jax2018github} and are thus automatically differentiable, enabling numerical evaluation of the Jacobian of the mapping from its forward implementation. 
A demonstration of this check is provided in the accompanying notebook.\footnote{\url{https://github.com/kemasuda/xv-reparam-paper/blob/main/analytic_jacobian.ipynb}.}

As shown in this example, automatic differentiation would provide a sufficiently efficient way to evaluate this Jacobian determinant in many practical cases. However, the analytic form is faster to compute and more convenient to implement, especially in routines that are not equipped with automatic differentiation. Thus, having the analytic Jacobian remains practically valuable regardless of whether automatic differentiation is available.

\section{Modeling of Lens Orbital Motion During Binary Microlensing Events}\label{sec:skowron}

The simplest model of a microlensing event caused by a binary lens assumes that the lens binary is fixed during the event. But sometimes the orbital motion of a binary --- namely changes in the sky-projected binary separation, rotation of the binary axis in the sky plane, and (to a subtler degree) motion along the line of sight --- has a detectable impact on the microlensing light curve. This ``lens orbital motion'' effect can provide additional constraints on the orbital parameters of microlensing exoplanets that are generally difficult to obtain, and also constrain the physical dimension of the lens system via its orbital timescale.

Since the orbital period of a lens binary as typically detected by microlensing is longer than the timescale of the microlensing event itself, the orbital motion, even if detected, usually covers only a small fraction of the entire orbit \citep[e.g.,][]{2010ApJ...713..837B}. This motivates a standard practice to model lens orbital motion using the instantaneous position $\x$ and velocity $\v$ at a reference epoch chosen during the microlensing event, using a coordinate frame whose $x$-axis is aligned with the sky-projected binary axis at the epoch and $z$-axis points toward the observer; see figure~8 of \citet{2011ApJ...738...87S}. The magnification due to microlensing depends on $\xv$ only through combinations normalized by the Einstein radius $\re$, and so the dimensionless parameters are introduced as \citep{2011ApJ...738...87S}:
\begin{align}
\label{eq:xv_ml}
    \x = 
    \re(s_0, 0, s_z),
    \quad 
    \v = \re s_0 \gammav.
\end{align}
The dimensionless positions $(s_0, s_z)$ and the velocity vector $\gammav$,
which has the dimension of (time)$^{-1}$, are used as the actual modeling parameters \citep[e.g., VBBinaryLensing code by][]{2021MNRAS.505..126B}.
While this completes the description of the motion in the binary-oriented frame, the microlensing signal also depends on the motion of the source star relative to the lens orbit. This orientation at a reference time is specified by the angle $\alpha_0$, again defined with respect to the projected binary axis ($x$-axis) at the reference epoch. This angle $\alpha_0$, in turn, fixes the orientation of the binary orbit relative to the lens--source proper motion --- so this degree of freedom corresponds to that of the longitude of the ascending node. 

In summary, standard fitting parameters of the binary geometry used for optimization and posterior sampling are $(s_0, s_z, \gammav, \alpha_0)$, while one often wants a prior PDF defined in the orbital element space. This is how the Jacobian determinant for the transformation between state vectors and orbital elements becomes relevant.

\subsection{Jacobian in \citet{2011ApJ...738...87S}}

For the conversion of the prior PDF, \citet{2011ApJ...738...87S} proposed to compute the determinant of the matrix
\begin{align}
\label{eq:skowron_jkep}
    j_{\rm kep}^{(\mathrm{S11})} \equiv {\partial(e, a, \tau, \Omega_{\rm node}, i,\omega) \over \partial(s_0, \alpha_0, s_z, \bm{\gamma})},
\end{align}
where $\omeganode$ is the longitude of the ascending node defined with respect to the {\it sky-projected binary axis at the reference epoch}, or $x$-axis in the {\it binary-oriented frame}. They then argue that
\begin{align}
\label{eq:skowron}
    \left|\det j_{\rm kep}^{(\mathrm{S11})}\right|^{-1}
    = \re^{-6} s_0^{-4}
    \left|\det\left({\partial(\x,\v) \over \partial(e,a,\tau,\omeganode,i,\omega)}\right)\right|
\end{align}
holds, without providing detailed reasoning. Although they proposed to compute the last determinant using LUP decomposition of the matrix whose elements are computed via the standard relations between $\xv$ and orbital elements, it can be computed analytically using Equation~\ref{eq:detjkep_atau} derived in Section~\ref{sec:detjkep}. Indeed, \citet{2011ApJ...738...87S} correctly notes that this Jacobian determinant goes to zero in proportion to $e\sin i$ --- Equation~\ref{eq:detjkep_atau} explicitly shows that this is the case for {\it any} values of $e$ and $\sin i$.

\subsection{Clarification of Their Derivation}

Equation~\ref{eq:skowron} is incorrect as written --- the determinant of $j_{\rm kep}^\mathrm{(S11)}$ defined by Equation~\ref{eq:skowron_jkep} is {\it identically zero}, because none of the elements in the numerator depends on $\alpha_0$. 
Here, it is essential to note that $\Omega_{\rm node}$ in this equation is defined with respect to the {\it sky-projected binary axis}; thus, all of the orbital elements in the numerator are determined once the orbit and the line-of-sight direction ($z$-axis) are specified, regardless of the direction of the projected binary axis ($x$-axis) with respect to the lens--source proper motion, as specified by $\alpha_0$. This can be made more explicit by writing down the orbital elements using $\x$ and $\v$ in Equation~\ref{eq:xv_ml}, as is also done in Appendix~B of \citet{2011ApJ...738...87S}: The components of the Lenz vector $\bm{e}$ and angular momentum vector $\bm{h}$ in the binary-oriented frame can all be specified using the components of $\x$ and $\v$ in the same frame, namely ($s_0$, $s_z$, $\bm{\gamma}$), without referring to $\alpha_0$. The same is true for the orbital elements referred to this frame, which are all scalars computed from the vectors $\bm{e}$ and $\bm{h}$.

To correctly incorporate the information of $\alpha_0$ into orbital elements, one needs to use $\Omega$ defined with respect to the {\it reference direction that is not tied to the binary orbit}, such as the direction of the source trajectory at the reference epoch to which $\alpha_0$ is referred, or the celestial north. If we choose the former --- as \citet{2011ApJ...738...87S} separates the proper motion vector $\bm{\mu}$ in the other block of the full Jacobian --- $\Omega$ is related to $\alpha_0$ as:
\begin{align}
    \label{eq:omegaalpha}
    \Omega = \Omega_{\rm node}(\elm') \pm \alpha_0 + \mathrm{const.},
\end{align}
where the argument of $\Omega_{\rm node}$ makes it explicit that this angle is determined by the other orbital elements $\elm'=(e,a,\tau,i,\omega)$ as in Section~\ref{ssec:detjkep5},
and the ``$\pm$'' sign and constant depend on the convention on the positive direction and reference axis of the angle $\alpha_0$.
Regardless of these conventions, Equation~\ref{eq:omegaalpha} yields
\begin{align}
    \left|\det\left({\partial(e, a, \tau, \Omega, i,\omega) \over \partial(e, a, \tau,\alpha_0, i,\omega)}\right)\right|
    = \left|\det \begin{pmatrix}
    \bm{1} & 0 \\
    {\partial \Omega_{\rm node} \over \partial \elm'} & \pm 1 \\
    \end{pmatrix}\right|
    = 1.
\end{align}
Thus, the corrected version of Equations~\ref{eq:skowron_jkep} and \ref{eq:skowron} evaluates to:
\begin{align}
\notag 
    \left|\det j_{\rm kep}^\mathrm{(corrected)}\right|^{-1} 
    &= \left|\det\left({\partial(s_0, \alpha_0, s_z, \bm{\gamma}) \over \partial(e, a, \tau, \Omega, i,\omega)}\right)\right|
    \quad (\text{using\ correct}\ \Omega)\\
    &= \left| \det\left({\partial(s_0, \alpha_0, s_z, \bm{\gamma}) \over \partial(e, a, \tau, \alpha_0, i,\omega)}\right) \right|
    = \left|\det\left({\partial(s_0, s_z, \bm{\gamma}) \over \partial(e, a, \tau, i,\omega)}\right)\right|
    = \re^{-5}\,s_0^{-3} \left|\det\left({\partial(x, z, v_x, v_y, v_z) \over \partial(e, a, \tau, i,\omega)}\right)\right|.
\end{align}
Using Equation~\ref{eq:detjkep5} to evaluate the last Jacobian determinant and substituting $x=\re s_0$ in the binary-oriented frame where $y=0$, we find
\begin{align}
    \left|\det j_{\rm kep}^\mathrm{(corrected)}\right|^{-1} 
    &= \re^{-5}s_0^{-3}\left|\det\jkep^{(a\tau)'}\right|=\re^{-5}s_0^{-3}\left|{1\over x}\det\jkep^{(a\tau)}\right|_{y=0} = \re^{-6}s_0^{-4}\left|\det\jkep^{(a\tau)}\right|.
\end{align}
The final result numerically agrees with equation B6 of \citet{2011ApJ...738...87S}.

In summary, the formula given by \citet{2011ApJ...738...87S} is incorrect as written, in that their $\Omega$ is defined with respect to the projected binary axis. When this error is corrected, their Jacobian determinant actually remains unchanged, because this determinant does not depend on the longitude of the ascending node in the first place, as shown in Section~\ref{sec:detjkep}.

\subsection{Note on the ``Uniform'' Prior for the Orbital Elements}\label{ssec:uniform}

The above formulation by \citet{2011ApJ...738...87S} implicitly assumes that the prior PDF for $\elm$ be defined using $\tau$ (rather than $M$), for which it is common and reasonable to adopt a uniform prior. This makes sense, but it may be worth noting that the normalization of the conditional prior $p(\tau|P)$ affects the implied marginal prior on $P$. Because $\tau$ is periodic, a sensible choice would be:
\begin{align}
\label{eq:ptauP}
    p(\tau|P) = \begin{cases}
        {1 \over P} \quad &\text{for}\quad \tau_{\rm ref} < \tau < \tau_{\rm ref} + P\\
        0 \quad &\text{otherwise}
    \end{cases},
\end{align}
where $\tau_{\rm ref}$ is an arbitrary reference time.
If one then chooses a prior PDF for $P$ as $\tilde p(P)$, the above choice ensures
\begin{align}
\label{eq:ptauP_marg}
    p(P) = \int p(\tau, P)\,\mathrm{d}\tau = \int p(\tau|P)\,\tilde p(P)\,\mathrm{d}\tau = \tilde p(P),
\end{align}
as one would intend. In this case, the joint prior PDF for $(\tau, P)$ does not have a uniform density in the ($\tau, P$)-plane even if $\tilde p(P)$ is uniform:
\begin{align}
\label{eq:ptauPjoint}
    p(\tau, P) = p(\tau|P)\,p(P) \propto {\tilde p(P) \over P}.
\end{align}

In some samplers like {\tt emcee} \citep{2013PASP..125..306F} in which one defines a function that returns the logarithm of the prior probability density, the prior in Equation~\ref{eq:ptauP} needs to be implemented as $\log(\mathrm{prior})=\mathrm{const.}-\log P$, rather than $\log(\mathrm{prior})=\mathrm{const.}$. 
If the ``$-\log P$'' term is omitted, the marginal PDF for $P$ is:
\begin{align}
    p(P) = \int p(\tau, P)\,\mathrm{d}\tau = \int p(\tau|P)\,p(P)\,\mathrm{d}\tau \propto P\cdot \tilde{p}(P).
\end{align}
This prior gives more weight at longer $P$ than Equation~\ref{eq:ptauP_marg} because the corresponding range of $\tau$ increases with $P$ when $p(\tau|P)$ is not properly normalized. This may not be the behavior one intends when specifying $\tilde{p}(P)$.

Equivalently, one can avoid this bookkeeping by using a phase variable normalized by the period, such as the mean anomaly at a reference epoch $M$, and assigning it a uniform prior \citep[e.g.,][]{2020AJ....159...89B}. When using $M$, for example, the resulting joint prior 
$p(M,P) = p(M|P)\,p(P) \propto \tilde p(P)$ yields $p(\tau,P)=p(M,P)\,|{\partial M\over \partial \tau}| \propto \tilde p(P)/P$, which is equivalent to Equation~\ref{eq:ptauPjoint}.

\newcommand{\amax}{a_\mathrm{max}}
\newcommand{\xmax}{x_\mathrm{max}}
\newcommand{\xmin}{x_\mathrm{min}}
\newcommand{\vmax}{v_\mathrm{max}}
\newcommand{\unif}{\mathcal{U}}
\newcommand{\norm}{\mathcal{N}}
\newcommand{\mass}{\mathcal{M}}
\newcommand{\dra}{\Delta\mathrm{RA}}
\newcommand{\ddec}{\Delta\mathrm{Dec}}

\section{Astrometry Fitting via Reparameterization using Cartesian State Vectors}\label{sec:astrometry}

Reparameterization using Cartesian state vectors can also be useful for orbit inference of directly imaged companions, particularly when the orbit is only partially observed and the constraints on the orbital elements are weak.
This idea has been explored in the literature \citep[e.g.,][]{2012A&A...542A..41C, 2016A&A...587A..89B}; in particular, \citet{2021RNAAS...5..162F} performed a performance comparison and reported improved convergence when sampling in state vectors. For a fair Bayesian comparison between parameterizations, however, one must ensure that the same prior density is used in both spaces, i.e., account explicitly for how the prior transforms under reparameterization. Otherwise the implied priors (and hence the posteriors) may differ between parameterizations, making the performance comparison not directly comparable.
We therefore revisit this comparison using matched priors in the two parameter spaces, in order to isolate the impact of the reparameterization. 
Here we employ a Markov Chain Monte Carlo (MCMC) algorithm that makes explicit use of the gradient of the posterior probability density, as it has demonstrated a promising improvement of the sampling performance compared to random-walk based MCMC methods \citep{2023AJ....166..164T}.

\begin{figure}
    \epsscale{1.0}
    \plotone{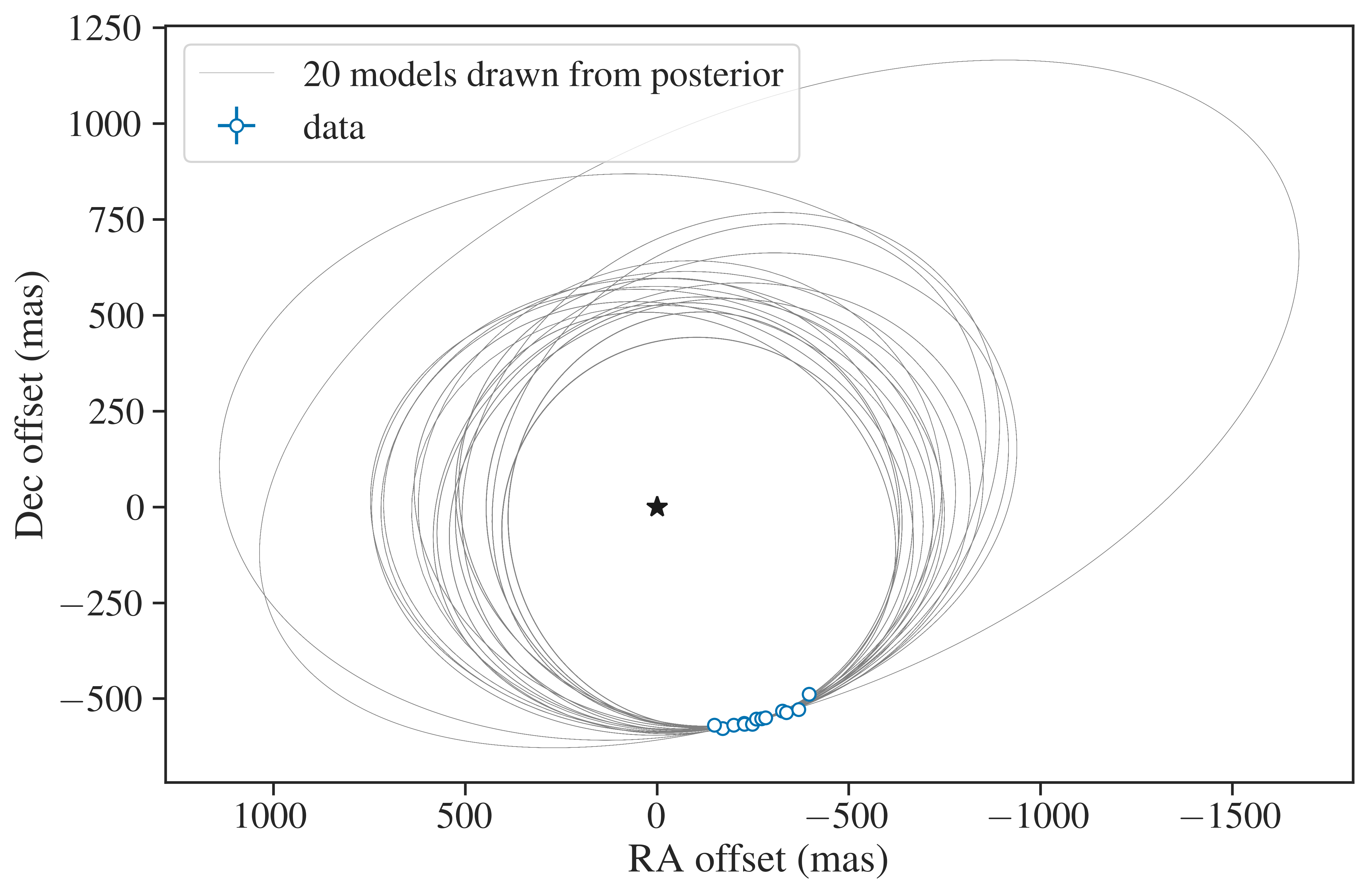}
    \caption{Our simulated data (circles) along with 20 models drawn from the posterior (gray lines). We choose the $+x$-axis in the positive Dec direction and $+y$-axis in the positive RA direction (left of this diagram). Thus the $+z$-axis points toward the observer.}
    \label{fig:models}
\end{figure}

We simulated astrometric observations similar to those for HR 8799d analyzed in \citet{2016AJ....152...28K}.
Specifically, we adopted the orbital elements in Table~2 of \citet{2019AJ....158....4O}, total mass of $1.5\,M_\odot$, and the parallax of $24.46\,\mathrm{mas}$ \citep{2023A&A...674A...1G} as the true input parameters.
Then we drew a single set of simulated observed values of the RA offsets $\dra_i$ and the Dec offsets $\ddec_i$ at the observed epochs labeled by the subscripts $i$ (circles in Figure~\ref{fig:models}), total mass $\mass$, and parallax $\varpi$, assuming independent Gaussian errors for these measurements with zero correlations. For the astrometric epochs and their errors ($\sigma_{\dra,i}$, $\sigma_{\ddec,i}$), we adopted the same values as the actual data in \citet{2016AJ....152...28K}. 
This choice resulted in an orbital phase coverage of $6\%$.
For the errors of the total mass and parallax, we adopted $\sigma_\mass=0.15\,M_\odot$ and $\sigma_\varpi=0.045\,\mathrm{mas}$, respectively. The resulting simulated set of observables, as well as the code used for fitting, are available from the author's GitHub.\footnote{\url{https://github.com/kemasuda/xv-reparam-paper/blob/main/fit_simulated_data.ipynb}}
Here we opted to use the simulated data so that the true parameters are known and the assumed likelihood (noise model) is correctly specified.

\begin{deluxetable}{l@{\hspace{1cm}}cc}[h!]
\caption{Priors adopted in the comparison in Section~\ref{sec:astrometry}.}
\label{tab:priors}
\tablehead{
	\colhead{Symbol} & \colhead{Meaning} & \colhead{Prior}
} 
\startdata
$P$ (days) & Orbital period & Log-uniform $(12705, 10^6)$\tablenotemark{$\dagger$}\\
$e$ & Orbital eccentricity & Uniform $(0, 0.99)$\\ 
$\cos i$ & Cosine of orbital inclination & Uniform $(-1, 1)$\\
$\Omega$ & Longitude of ascending node & Uniform $(-\pi, \pi)$\\ 
$\omega$ & Argument of periapsis & Uniform $(-\pi, \pi)$\\ 
$M$ & Mean anomaly at the reference epoch & Uniform $(-\pi, \pi)$
\enddata
\tablenotetext{\dagger}{Minimum value was chosen to be five times the data duration.}
\end{deluxetable}

We then sampled from the joint posterior distribution for the orbital elements $\elm$, mass, and parallax, conditioned on this simulated data set $\mathcal{D}$, $p(\elm,\mass,\varpi|\mathcal{D}) \propto \mathcal{L}(\elm,\mass,\varpi)\,\pi(\elm,\mass,\varpi)$ where $\mathcal{L}$ and $\pi$ denote the likelihood function and the prior PDF, respectively, with and without reparameterization to Cartesian vectors. We adopted the log-likelihood function given by
\begin{align}
\notag
    \log \mathcal{L}(\elm,\mass,\varpi) 
    &= -{1\over 2}\sum_i \left\{\left[(\hat{\ddec}_i - \varpi\cdot x_i(\elm,\mass)) \over \sigma_{\ddec,i}\right]^2
    + \left[(\hat{\dra}_i - \varpi\cdot y_i(\elm,\mass)) \over \sigma_{\dra,i}\right]^2\right\} \\
    &\quad - {1\over 2}\left({\hat{\mass}-\mass \over \sigma_\mass} \right)^2 - {1\over 2}\left({\hat{\varpi}-\varpi \over \sigma_\varpi} \right)^2 + \mathrm{const.},
\end{align}
where symbols with hats denote simulated observables and those without denote either sampled parameters ($\mass$, $\varpi$) or their deterministic functions ($x_i$ and $y_i$); here $x_i$ and $y_i$ are physical offsets of the companion relative to the central star in the {\it Dec and RA directions}, respectively, which are converted to angular offsets by multiplying by the parallax $\varpi$. The physical coordinates were computed from the orbital elements following a standard procedure \citep[e.g.,][]{1999ssd..book.....M}, where the Kepler equation was solved following \citet{1995CeMDA..63..101M}.
We adopted the priors as listed in Table~\ref{tab:priors}, which are uniform or log-uniform in the orbital elements, but performed this sampling in two different ways: (i) sample the orbital elements directly from the prior in Table~\ref{tab:priors}, and (ii) sample the state vectors at a reference epoch (chosen to be the time of the first observation), $(\x, \v)$, from a uniform distribution as described in Section~\ref{ssec:sample_xv}, and then apply the log-probability correction using the Jacobian determinant as in Equation~\ref{eq:pxv_from_pelm}, so that the prior PDF induced for the orbital elements remains essentially the same as adopted in (i).
We implemented these conversion and sampling using JAX \citep{jax2018github} as part of the {\tt jnkepler} package \citep{2024AJ....168..294M, 2025ascl.soft05006M}, so that the derivatives are available and can be used for MCMC sampling. Specifically, we used the No-U-Turn Sampler \citep[NUTS;][]{2011arXiv1111.4246H}, a variant of Hamiltonian Monte Carlo \citep{2012arXiv1206.1901N}, as implemented in {\tt NumPyro} \citep{phan2019composable}. 
For this test, we ran four independent chains for 7,500 warm-up steps and for 7,500 sampling steps to obtain 30,000 samples in total.
We adopted the target acceptance probability of $0.99$, and set the maximum tree depth for NUTS to be 13. Here we made these more conservative choices than the default values in the {\tt NumPyro} implementation, as we found that the non-Gaussianity in the posterior causes a large number of divergent transitions (a warning that the numerical integration used to propose a new state became unstable and that the resulting sample may be unreliable) with the default settings, even in the reparameterized case.
In both runs, we achieved split Gelman--Rubin statistics \citep{BB13945229} of $<1.01$ in $\sim 10\,\mathrm{min}$ using four Intel Core i7 CPUs (3.8~GHz).

\begin{figure*}
    \epsscale{1.1}
    \plotone{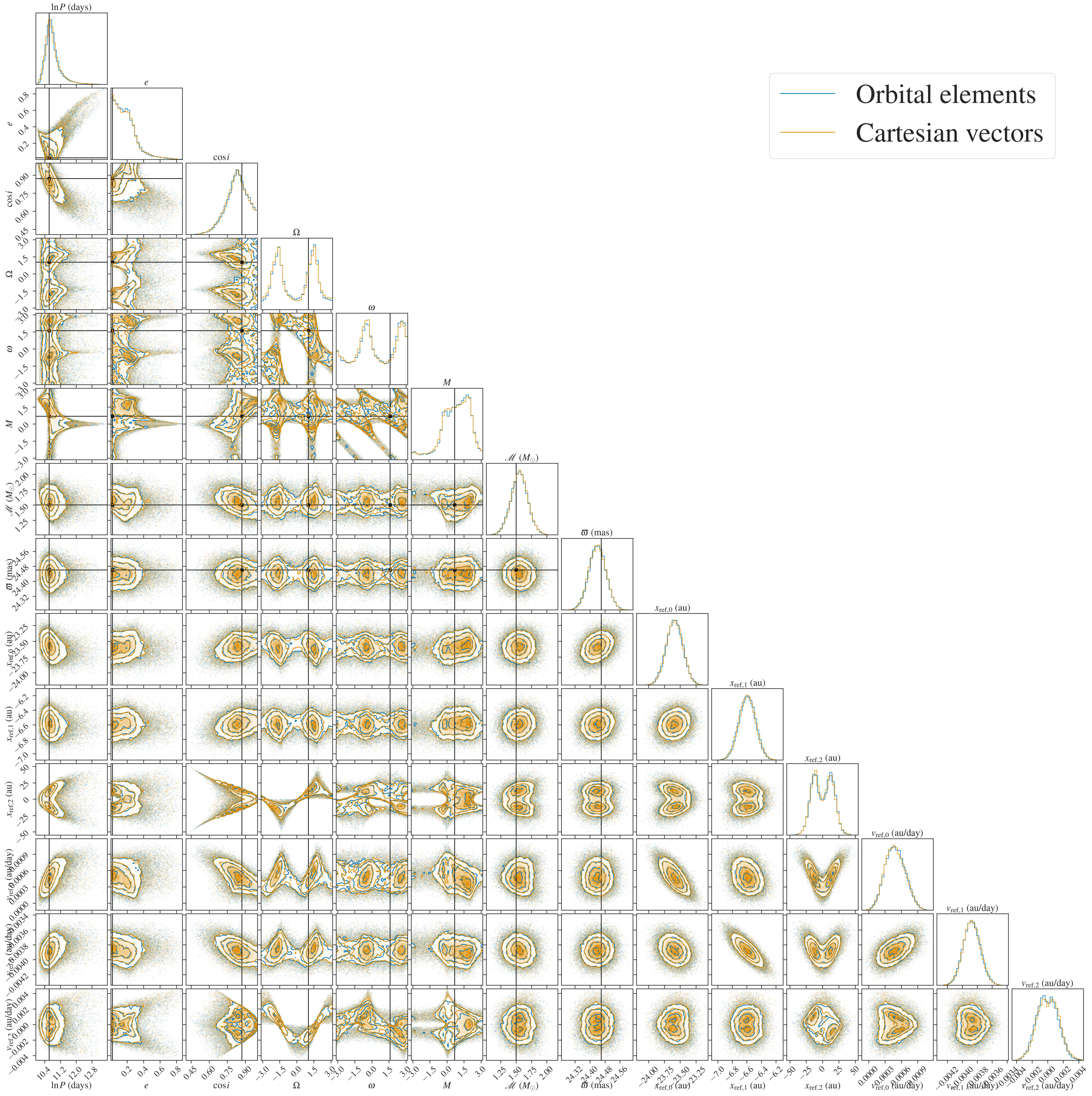}
    \caption{Corner plot comparing posterior samples obtained by directly sampling the
    orbital elements (blue) and by sampling the Cartesian state vectors at the
    reference epoch (orange). The same physical prior is imposed in both
    parameterizations through the Jacobian transformation, so the two posteriors
    overlap as expected. Samples are shown in both orbital-element coordinates 
    (first six entries) and Cartesian phase-space coordinates (last six entries).
    Black lines indicate the input values used to generate the simulated data.}
    \label{fig:corner}
\end{figure*}

Figure~\ref{fig:corner} shows a corner plot for the posterior samples obtained by directly sampling from the orbital elements (blue) and via reparameterization to the Cartesian state vectors (orange). We recover the identical distributions, confirming that our prior conversion has been performed correctly. They are also consistent with the input values (black solid lines), as expected.
We also note complicated patterns of correlations in the orbital elements (first six entries), which is mitigated in the state vectors (last six entries). This feature makes sampling easier in this basis. In Figure~\ref{fig:models}, the models drawn from the posterior are overlaid with the data (gray lines).

\begin{figure}
    \epsscale{1.0}
    \plotone{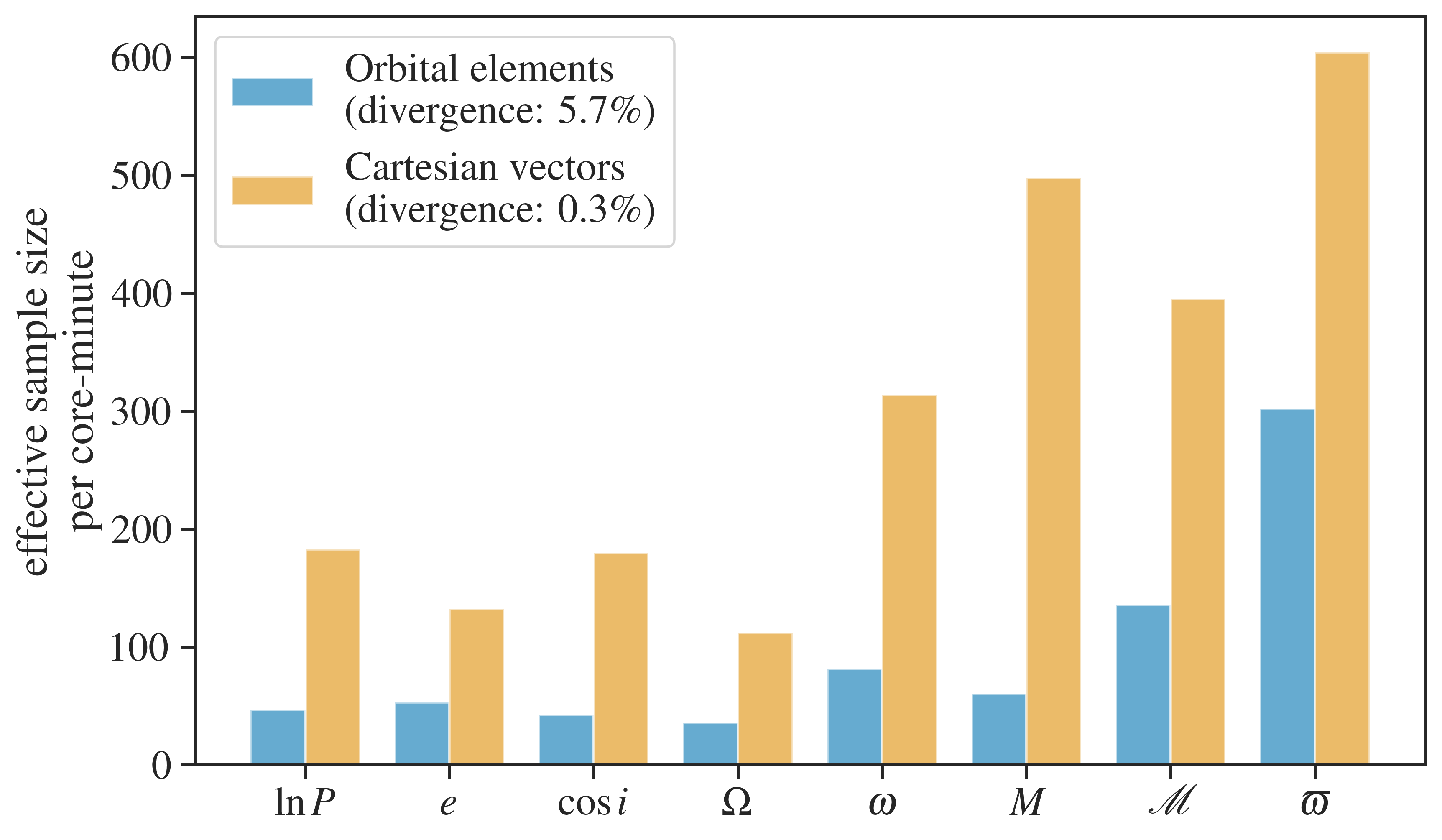}
    \caption{The effective sample size (ESS) per core-minute for the selected parameters sampled with (orange) and without (blue) reparameterizing to Cartesian state vectors.}
    \label{fig:ess}
\end{figure}

Following \citet{2023AJ....166..164T}, we compare the sampling efficiency of the two parameterizations primarily using the effective sample size (ESS) per core-minute based on four MCMC chains (i.e., total ESS divided by (number of chains $\times$ the wall-clock time)), which were computed using \texttt{arviz.summary} in ArviZ \citep{2019JOSS....4.1143K, 2021BayAn..16..667V}.
Because the two approaches operate in different parameter spaces, we report ESS for a common set of derived orbital elements, which are typically the primary quantities of interest in orbit inference.
The results are shown in Figure~\ref{fig:ess}. Across the parameters considered, sampling in Cartesian state vectors yields an improvement of a factor of $2$--$7$ in ESS per core-minute. We also find that the fraction of divergent transitions (see legends) is reduced when using Cartesian vectors, suggesting a posterior geometry that is easier for Hamiltonian Monte Carlo to explore under this reparameterization.
The absolute values of ESS and divergence fractions depend on the specific problem, hardware, and tuning choices (e.g., target acceptance probability and maximum tree depth). Nevertheless, we find qualitatively consistent improvements across a range of configurations when adopting the Cartesian state-vector parameterization.

We emphasize that this benchmark is intended as a relative comparison between the two parameterizations, rather than an attempt to identify the most efficient inference strategy for this particular example. In this low-dimensional case, the prior and posterior overlap substantially, so rejection sampling (or closely related methods) as implemented in, e.g., {\tt orbitize!} \citep[OFTI;][]{2017AJ....153..229B, 2020AJ....159...89B} is likely more efficient in practice.
However, such approaches can become challenging in higher-dimensional settings with many free parameters --- for example, when fitting multiple planets and/or combining multiple data sets. The microlensing analysis discussed in Section~\ref{sec:skowron} provides such an example, where orbital inference is performed jointly with additional microlensing parameters.
In these regimes, reparameterization to Cartesian state vectors may offer a useful and broadly applicable alternative.

\subsection{Sampling from Cartesian State Vectors}\label{ssec:sample_xv}

Here we describe our method of sampling $\x$ and $\v$ from bound Keplerian orbits. In principle, for a random-walk Metropolis--Hastings sampler one could simply enforce the physical constraint by rejecting proposals with positive orbital energy. However, this approach can become inefficient in practice, as a substantial fraction of random-walk proposals fall into the unbound region and are rejected near the boundary, leading to poor mixing. We therefore construct a sampling scheme that generates proposals directly within the bound region. The same idea may also be useful for rejection-based samplers, where proposing only physically allowed orbits reduces wasted draws.

Given the total mass $\mu=\mathcal{G}\mass$, we would like to sample $\x$ and $\v$ at a reference epoch from a uniform distribution whose support is the region of the $(\x,\v)$ space corresponding to a bound orbit (i.e., negative total energy).
We implement this sampling in the following two steps: (i) first sample $\x$ from a uniform distribution, and then (ii) sample $\v$ uniformly so that the total energy remains negative, given the value of $\x$ sampled in the first step.

\paragraph{Uniform Sampling of $\bm{x}$} 

Given the total mass, the maximum orbital period translates into $\amax$, the maximum possible semi-major axis. Given $\amax$, $|\x|$ must satisfy $|\x| < \xmax = 2\amax$. We thus sample isotropic $\x$ that satisfies $\xmin < |\x| < \xmax$. The PDF for $|\x|$ in this case is:
\begin{align}
    p(|\x|) = {|\x|^2 \over {\xmax^3/3 - \xmin^3/3}},
\end{align}
whose CDF $F(|\x|)$ for $\xmin < |\x| < \xmax$ is
\begin{align}
    F(|\x|) = {{|\x|^3 - \xmin^3} \over {\xmax^3 - \xmin^3}}.
\end{align}
So $|\x|$ can be sampled by sampling $u_x \sim \unif(0,1)$ and by computing $|\x|=\left[\xmin^3 + u_x(\xmax^3 -\xmin^3)\right]^{1/3}$. To improve the numerical stability, here we also set a non-zero lower limit $\xmin$, which is chosen sufficiently small compared to the range allowed by the actual data. Note that $\xmin$ cannot be simply related to either of minimum semi-major axis or maximum eccentricity. 

This $|\x|$ can then be multiplied by an isotropic 3D vector $\hat{\x}$ whose norm is unity. This $\hat{\x}$ is obtained by sampling each Cartesian component from independent normal distributions, $x_i \sim \norm(0, 1)$ ($i=1,2,3$), and then by computing $\hat{\x}=(x_1,x_2,x_3)/\sqrt{x_1^2+x_2^2+x_3^2}$. The full vector $\x$ is then given as $\x = |\x|\hat{\x}$.

\paragraph{Uniform Sampling of $\bm{v}$ Given $\bm{x}$}

Given $\x$, the norm of $\v$ must satisfy
\begin{align}
    |\v| < \sqrt{\mu\left( {2\over|\x|} - {1\over{\amax}}\right)} \equiv \vmax(x)
\end{align}
so that the orbit is bound and has $a<\amax$. The same algorithm as we used to sample $\x$ allows us to sample from the PDF
\begin{align}
\label{eq:pvgivenx}
    p(\v | \x) = {1 \over {4\pi\over 3}\vmax(x)^3}, \quad 0 < |\v| < \vmax(x).
\end{align}
However, this way we are not sampling from a uniform distribution for $(\x,\v)$, because the normalization of $p(\v|\x)$ in Equation~\ref{eq:pvgivenx} depends on $\x$: the joint PDF we are actually drawing from is
\begin{align}
    p(\x, \v) = p(\v|\x)\,p(\x) \propto {1 \over {4\pi\over 3}\vmax(x)^3},
\end{align}
whose density is not uniform in the $\xv$-space.\footnote{This situation is analogous to what was discussed in Section~\ref{ssec:uniform}.} We therefore apply the log-probability correction $3\log \vmax(x)$ after the above procedure, so that the resulting sampling is uniform over the bounded region in the $\xv$-space.

\subsubsection{Sampling via the Orbital Period}

Sometimes additional data such as radial velocities may provide a better constraint on the orbital period $P$. Even with partial astrometry alone, the presence of other planets may provide information on the possible range of orbital semi-major axis (and hence orbital period). Including $P$ as a sampling parameter may make it easier to incorporate such information.

This can be done by replacing one component of the Cartesian state vectors with $P$. One simple choice is to replace $|\v|$. Given $P$, the Kepler's third law fixes the semi-major axis $a_P$, and hence the specific orbital energy $(1/2)|\v|^2 - \mu/|\x|=-\mu/(2a_P)$. Thus, given total mass, $P$, and $|\x|$, the norm of the velocity vector is uniquely determined as
\begin{align}
    |\v|_P = \sqrt{\mu\left({2\over|\x|} - {1\over a_P}\right)}.
\end{align}
Therefore, the sampling via $P$ can be performed by (i) sampling $\x$ as above, (ii) sampling $P$ from a certain prior PDF, and then (iii) sampling $\hat{\v}$ as above and by computing $\v=|\v|_P\,\hat{\v}$. Again, a caveat here is that this way we are sampling $(P,\hat{\v})$ instead of $(|\v|, \hat{\v})$, and so the additional Jacobian correction for this transformation is needed. The Jacobian determinant corresponding to this transformation is:
\begin{align}
    \left|\det\left({\partial \v \over {\partial P\partial \hat{\v}}}\right)\right| 
    = |\v|_P^2\left|{\partial |\v|_P \over \partial P}\right| = \mu {|\v|_P^2 \over 2|\v|_Pa^2}{2\over 3}{a\over P}
    ={\mu |\v|_P \over 3a_P\,P}.
\end{align}

\section{Summary}

We have derived an analytic, closed-form expression for the Jacobian determinant of the transformation between Cartesian phase-space coordinates and the standard orbital elements in the Kepler problem. The resulting formula,
\begin{align}
\notag 
\det\left({\partial(\x,\v)\over\partial(a,e,i,\Omega,\omega,M)}
\right) = \frac{1}{2}\,\mu^{3/2}\, a^{1/2}\, e\, \sin i,
\end{align}
is compact and straightforward to implement, and it enables transparent conversion of prior densities between the two parameterizations for Bayesian orbit inference without resorting to numerical Jacobians or automatic differentiation.

We illustrate the utility of this result in two settings. First, we clarify the widely used microlensing formulation of lens orbital motion by \citet{2011ApJ...738...87S}. We show that their definition of the nodal angle unintentionally ties the longitude of the ascending node to the sky-projected binary axis, rendering the intermediate Jacobian formally singular. When the nodal angle is instead defined with respect to an axis independent of the binary orbit, the corrected Jacobian is well-defined and practically remains unchanged, because $\det\jkep$ carries no dependence on the longitude of ascending node.

Second, we quantify the impact of the same reparameterization on astrometric orbit fitting with a gradient-based MCMC sampler (NUTS). We find that reparameterizing to Cartesian state vectors improve the robustness and efficiency of the sampling over sampling in orbital elements. This provides a quantitative underpinning for the common but implicit expectation --- also reflected in microlensing practice --- that state-vector parameterizations can be advantageous in weakly constrained orbit inference.

\begin{acknowledgments}

The authors thank Eiichiro Kokubo for helpful discussions that led us to relevant references on this topic. We also thank the anonymous reviewer for a careful reading of the manuscript and for thoughtful comments that helped improve it.
Work by KM was supported by JSPS KAKENHI grant Nos.~21H04998 and 25K07387.
KN was supported by the JST Next Generation Researcher Challenging Research Program at Osaka University.

\end{acknowledgments}

\software{ArviZ \citep{2019JOSS....4.1143K}, corner.py \citep{corner}, JAX \citep{jax2018github}, jnkepler \citep{2025ascl.soft05006M}, NumPyro \citep{phan2019composable}}




\bibliography{references_masuda}
\bibliographystyle{aasjournalv7}



\end{document}